\documentclass[]{aastex631}
\usepackage{amsmath}

\begin{document}

\title{Photometric and Spectroscopic Studies of Two Early-Type Eclipsing Binaries}

\author{Pengfei Ye}
\affiliation{Shandong Key Laboratory of Optical Astronomy and Solar-Terrestrial Environment, 
School of Space Science and Technology, Institute of Space Sciences, Shandong University, Weihai, Shandong 264209, People's Republic of China; \href{mailto:kaili@sdu.edu.cn}{kaili@sdu.edu.cn}}

\author[0000-0003-3590-335X]{Kai Li}
\affiliation{Shandong Key Laboratory of Optical Astronomy and Solar-Terrestrial Environment, 
School of Space Science and Technology, Institute of Space Sciences, Shandong University, Weihai, Shandong 264209, People's Republic of China; \href{mailto:kaili@sdu.edu.cn}{kaili@sdu.edu.cn}}

\author{Fei Liu}
\affiliation{Shandong Key Laboratory of Optical Astronomy and Solar-Terrestrial Environment, 
School of Space Science and Technology, Institute of Space Sciences, Shandong University, Weihai, Shandong 264209, People's Republic of China; \href{mailto:kaili@sdu.edu.cn}{kaili@sdu.edu.cn}}

\author[0009-0009-6364-0391]{Xiang Gao}
\affiliation{Shandong Key Laboratory of Optical Astronomy and Solar-Terrestrial Environment, 
School of Space Science and Technology, Institute of Space Sciences, Shandong University, Weihai, Shandong 264209, People's Republic of China; \href{mailto:kaili@sdu.edu.cn}{kaili@sdu.edu.cn}}

\author{Jing-Yi Wang}
\affiliation{Shandong Key Laboratory of Optical Astronomy and Solar-Terrestrial Environment, 
School of Space Science and Technology, Institute of Space Sciences, Shandong University, Weihai, Shandong 264209, People's Republic of China; \href{mailto:kaili@sdu.edu.cn}{kaili@sdu.edu.cn}}

\begin{abstract}

Using high-precision photometric data from TESS, medium resolution spectroscopic data from LAMOST, and long-term eclipse timings, we provide orbital parameters for two early-type detached eclipsing binary systems: TYC 3740-2072-1 and TYC 2888-780-1, and analyze the orbital period variations and evolutionary status of these two targets. TYC 3740-2072-1, with a spectral type of B1V, consists of a $6.914 M_\odot$ subgiant and a $6.233 M_\odot$ main-sequence component. It is expected to evolve into a semi-detached binary, potentially serving as a progenitor of a Type Ia supernova. TYC 2888-780-1 has a spectral type of A3 and consists of two main-sequence components with masses of $1.682 M_\odot$ and $1.673 M_\odot$, respectively. It is undergoing stable evolution. Through eclipse timing analysis, we find that both targets exhibit apsidal motion effects, with observed AM rates of $\dot{\omega}_{obs} = 0.0412~\text{deg}~\text{cycle}^{-1}$ and $\dot{\omega}_{obs} = 0.0205~\text{deg}~\text{cycle}^{-1}$, respectively. Additionally, TYC 2888-780-1 exhibits orbital period variations, which we attribute to the light travel time effect caused by a third body, the minimum mass of this third body is estimated to be $0.598 M_\odot$.

\end{abstract}


\section{Introduction} \label{sec:1}

Early-type stars are those with spectral types O, B, and A. They are characterized by high mass, high temperature, and high luminosity. The formation conditions of early-type binaries are complex. Compared to low-mass stars, they are less numerous, evolve more rapidly, and have shorter lifetimes. Observations indicate that the binary fraction among early-type stars is very high \citep{2012Sci...337..444S, 2022A&A...667A..44G}, and one component of a binary system significantly influences the evolution of its companion, leading to substantial differences from single-star evolution \citep{2009AJ....137.3358M}. These binary systems are considered as progenitors of supernovae and compact objects such as neutron stars, and black holes, and they are also the origins of various astrophysical phenomena, including gravitational wave events, gamma-ray bursts, and X-ray binaries \citep{2002ApJ...572..407B, 2012Sci...337..444S}.

Therefore, the observation and study of early-type binary stars are of significant importance in astrophysics. Accurate determination of fundamental physical parameters such as stellar mass, radius, and luminosity is crucial for understanding the structure of early-type stars and calibrating stellar evolutionary models \citep{1991A&ARv...3...91A}. This study aims to detect early-type eclipsing binaries and obtain their precise physical parameters, providing essential data for studying the progenitors of compact objects.

TYC 3740-2072-1 was identified as a spectroscopic binary by \citet{2003A&A...406..119N}, with a spectral type of B1V. Recently, the Gaia survey confirmed it as an eclipsing binary \citep{2022yCat.1358....0G}; TYC 2888-780-1 was identified as an eclipsing binary by \citet{2018ApJS..235...41K}, with a spectral type of A3. The Gaia DR3 mission \citep{2023A&A...674A..14R} provided atmospheric parameters for both targets. The basic information of the two targets is summarized in Table \ref{table 1}.

The orbital periods of TYC 3740-2072-1 and TYC 2888-780-1 are 3.8669139 days \citep{2021MNRAS.503..200J} and 1.4547185 days \citet{2018ApJS..235...41K}, respectively. Complete light curves are difficult to obtain for such long period binaries. The Transiting Exoplanet Survey Satellite (TESS) \citep{2015JATIS...1a4003R}, with its long baseline, short exposure times, and high-precision photometric measurements, has provided high-quality observational data for these two targets. Additionally, the Large Sky Area Multi-Object Fiber Spectroscopy Telescope (LAMOST) \citep{2012RAA....12.1197C, 2015RAA....15.1095L} has provided multiple medium-resolution spectroscopic observations for the two targets. By combining these photometric and spectroscopic data, we can derive the orbital parameters of these two targets and conduct further investigations.

Section 2 introduces the observational data used in this work. Section 3 describes the extraction and adjustment of radial velocities. In Section 4, we determine the orbital parameters of the two targets using the W-D code. Section 5 presents the eclipse timing analysis of the two targets. Finally, discussions and conclusions are shown in Section 6.

\begin{deluxetable}{lcccc}
\setlength{\tabcolsep}{15pt}
\tablenum{1}
\tablecaption{The Information of the Two Systems
\label{table 1}}
\tablewidth{0pt}
\tablehead{
\colhead{Name} & \colhead{Other Name} & \colhead{$T_{\mathrm{eff, 1}}$} & \colhead{Period} & \colhead{TESS Mag} \\
\colhead{} & \colhead{} & \colhead{(K)} & \colhead{(days)} & \colhead{}
}
\startdata
TYC 3740-2072-1 & LS V +56 59 & 18147.3 & 3.8669139 & 10.338 \\
TYC 2888-780-1  & HD 276600   & 10902.5 & 1.4547185 & 10.463 \\
\enddata
\end{deluxetable}

\section{Observed data} \label{sec:2}

\subsection{Photometric observation} \label{sec:2.1}

TESS is a full-sky survey mission for transiting exoplanets, led by the Massachusetts Institute of Technology (MIT) and funded by NASA \citep{2015JATIS...1a4003R}. The primary goal of TESS is to detect Earth-sized planets orbiting bright stars, enabling follow-up observations to determine their mass, atmospheric composition, and other properties. TESS consists of four high-precision, red-end-sensitive wide-field cameras, each covering a field of view of $ 24^{\circ} \times 96^{\circ} $ in the sky. The detector operates in a wavelength range of 600-1000 nm. The excellent quality of TESS data makes it suitable for studying eclipsing binary events \citep{2021A&A...649A..64B}. However, mission-specific planning remains essential for observations of stellar targets. In the future, multi-band space-based photometric surveys tailored to stellar astrophysics will provide more accurate temperature determinations and broader stellar characterization capabilities.

TESS observed TYC 3740-2072-1 three times in sectors 19, 59, and 73, with an exposure time of 1800s in sector 19 and 200s in the other sectors. Similarly, TESS observed TYC 2888-780-1 twice in sectors 19 and 59, with exposure times of 1800s and 200s, respectively. We downloaded the TESS light curve data for these two binary stars from the MAST database, processed by the MIT Quick-Look Pipeline (QLP) \citep{2020RNAAS...4..204H, 2020RNAAS...4..206H}, and used the Simple Aperture Photometry (SAP) data from the SPOC data reduction pipeline and converted the normalized flux values into magnitude using the following equation:

\begin{equation}
    m=m_\mathrm{TESS} - 2.5 log(flux_\mathrm{SAP})
\end{equation}

\begin{figure*}
\includegraphics[width=\textwidth]{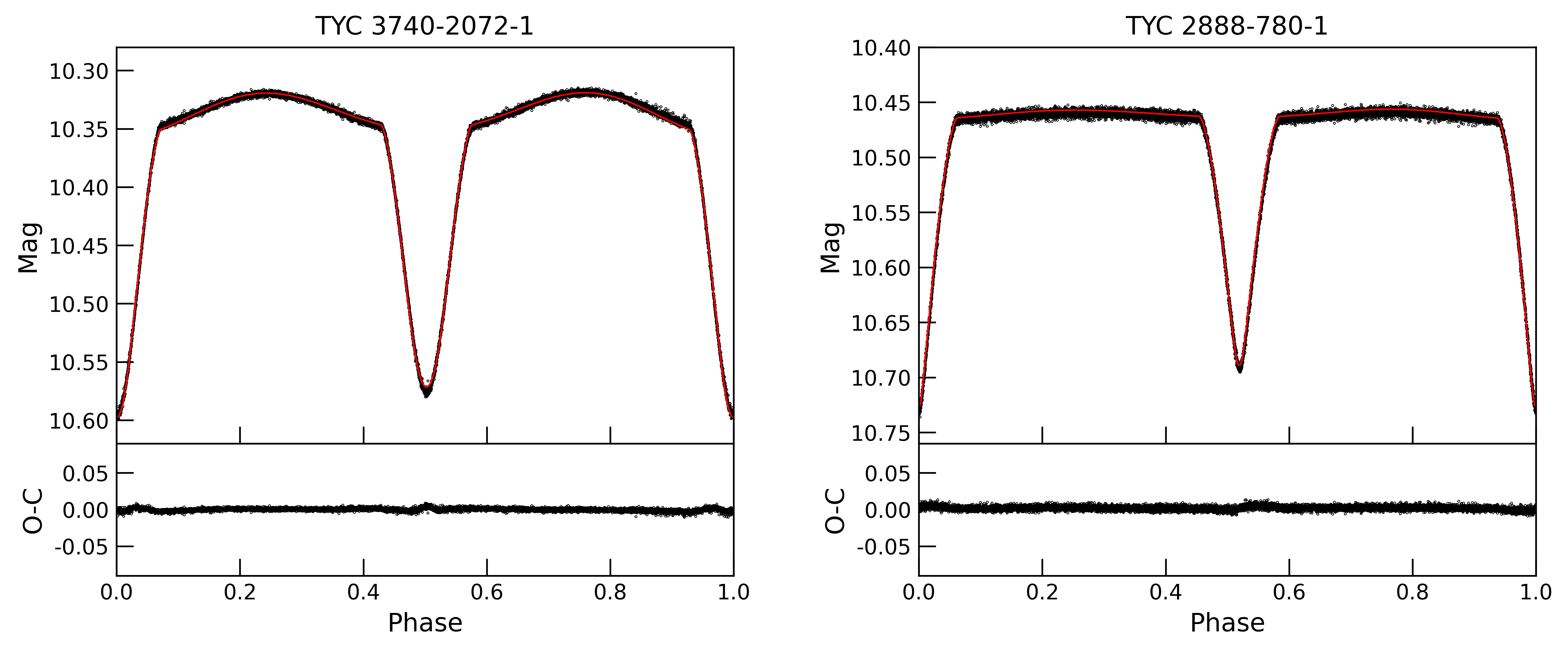}
\caption{Light curves of the two systems. The red solid line in the upper part is the theoretical light curve, the black circle is the actual observed light curve, and the lower part shows the difference between the observed value and the theoretical curve.
\label{fig:1}}
\end{figure*}

Using the Lomb-Scargle periodogram method \citep{2009A&A...496..577Z}, we derive a period of 3.8672357 days for TYC 3740-2072-1, which is close to the period of 3.8669139 days reported by ASAS-SN \citep{2021MNRAS.503..200J}. Similarly, we determined the period of TYC 2888-780-1 to be 1.4547250 days, which is in close agreement with the value of 1.4547185 days given by \citet{2018ApJS..235...41K}. The folded light curves are shown in Figure \ref{fig:1}.

\subsection{Spectral observation} \label{sec:2.2}

LAMOST is a new type of spectroscopic survey telescope that combines both a wide field of view and a large aperture \citep{2012RAA....12.1197C, 2015RAA....15.1095L}. It is equipped with 4000 fibers and has a field of view of 5$^\circ$. LAMOST is located at the Xinglong Observing Station of the National Astronomical Observatories (117$^\circ$34$'$30$''$E, 40$^\circ$23$'$36$''$N) at an altitude of 960 meters. The medium-resolution spectroscopic survey (MRS) of LAMOST has a resolution of R \verb|~| 7500. The wavelength coverage for the blue band (B) is 4950 - 5350 \AA, and for the red band (R), it is 6300 - 6800 \AA. \citep{2012RAA....12..723Z}

The two target binaries were observed using the medium-resolution mode of LAMOST. Since both targets have relatively high temperatures, spectra with blue-end exposure were selected. A total of 35 high-quality (signal-to-noise ratio $>$ 10) medium-resolution spectra were obtained, including 24 spectra for TYC 3740-2072-1 and 11 spectra for TYC 2888-780-1. The observational information is summarized in Table \ref{table 2}.

\section{Radial velocity determination} \label{sec:3}

Based on the atmospheric information of the two targets, for TYC 2888-780-1, we use the local thermodynamic equilibrium (LTE) PHOENIX model \citep{2013A&A...553A...6H} to calculate synthetic spectra. Since the Gaia temperature of TYC 3740-2072-1 exceeds the limit of the PHOENIX model, 12000K, we instead decided to use the non-local thermodynamic equilibrium (N-LTE) CMFGEN model \citep{1998ApJ...496..407H,2010A&A...516A..13P} to calculate the synthetic spectrum. Using the synthetic spectrum as template spectrum, we use the cross-correlation function (CCF) method \citep{2007A&A...465..943S,2010AJ....140..184M} to compute the CCF between the template spectrum and the observed spectrum,

\begin{equation}
CCF(v)=\int S(\lambda) \cdot T(\lambda,v) d\lambda,
\end{equation}
where $v$ represents the radial velocity in $km~s^{-1}$, $\lambda$ is wavelength in \AA, $S(\lambda)$ denotes the observed spectrum, and $T(\lambda,v)$ refers to the template spectrum. The computed $\text{CCF}(v)$ is fitted with Gaussian components using GaussPy+ \citep{2019RAA....19...99M}. For phases close to 0 and 0.5, $\text{CCF}(v)$ contains only a single Gaussian component, while a double Gaussian fitting is used for other cases. The fitted center corresponds to the radial velocity of each stellar component. The calculation results are presented in Table \ref{table 2}.

\begin{deluxetable*}{lccccc}
\setlength{\tabcolsep}{10pt}
\tablenum{2}
\tablecaption{The Radial Velocity Values of Two Systems
\label{table 2}}
\tablewidth{0pt}
\tablehead{
\colhead{Name} & \colhead{BJD} & \colhead{Phase} & \colhead{$\mathbf{S/N}$} & \colhead{$\mathbf{RV_1}$} & \colhead{$\mathbf{RV_2}$} \\
\colhead{} & \colhead{($+$2400000)} & \colhead{} & \colhead{} & \colhead{($km/s$)} & \colhead{($km/s$)}
}
\startdata
TYC 3740-2072-1 & 
     58795.22152 & 0.000 & 55  & -                 &  -59.0 $\pm$ 3.1 \\
{} & 58478.11337 & 0.001 & 75  & -                 &  -52.0 $\pm$ 2.5 \\
{} & 58795.23775 & 0.004 & 62  & -                 &  -60.4 $\pm$ 2.4 \\
{} & 58478.13401 & 0.007 & 66  & -                 &  -52.9 $\pm$ 2.9 \\
{} & 58795.25398 & 0.008 & 68  & -                 &  -60.9 $\pm$ 3.1 \\
{} & 58478.15028 & 0.011 & 66  & -                 &  -49.9 $\pm$ 2.6 \\
{} & 58478.16652 & 0.015 & 68  & -                 &  -56.4 $\pm$ 3.0 \\
{} & 58850.04889 & 0.177 & 81  & -189.7 $\pm$ 14.0 &   89.3 $\pm$ 7.4 \\
{} & 58850.06510 & 0.182 & 79  & -197.8 $\pm$ 16.0 &   89.4 $\pm$ 7.8 \\
{} & 58850.08134 & 0.186 & 72  & -195.3 $\pm$ 13.7 &   95.5 $\pm$ 5.9 \\
{} & 58850.09757 & 0.190 & 77  & -218.2 $\pm$ 12.6 &   92.8 $\pm$ 6.0 \\
{} & 58850.11376 & 0.194 & 80  & -199.6 $\pm$ 17.8 &   92.5 $\pm$ 7.8 \\
{} & 58850.13000 & 0.198 & 79  & -216.4 $\pm$ 12.2 &  100.0 $\pm$ 5.3 \\
{} & 58850.14620 & 0.203 & 77  & -203.2 $\pm$ 13.8 &  101.0 $\pm$ 5.9 \\
{} & 58831.07832 & 0.272 & 100 & -225.2 $\pm$ 10.6 &   97.7 $\pm$ 4.7 \\
{} & 58831.09455 & 0.276 & 96  & -221.5 $\pm$ 10.0 &   99.6 $\pm$ 5.1 \\
{} & 58831.11085 & 0.280 & 99  & -219.0 $\pm$ 11.7 &  104.8 $\pm$ 5.5 \\
{} & 58481.07975 & 0.768 & 55  &   92.2 $\pm$ 10.5 & -228.2 $\pm$ 4.1 \\
{} & 58481.09594 & 0.773 & 53  &   99.9 $\pm$ 12.4 & -222.9 $\pm$ 6.5 \\
{} & 58481.11213 & 0.777 & 55  &  107.9 $\pm$ 10.6 & -224.2 $\pm$ 4.6 \\
{} & 58481.12832 & 0.781 & 55  &  107.0 $\pm$  8.7 & -221.6 $\pm$ 3.3 \\
{} & 58481.14452 & 0.785 & 52  &  110.1 $\pm$ 11.5 & -216.4 $\pm$ 5.8 \\
{} & 58478.08054 & 0.993 & 69  & -                 &  -61.4 $\pm$ 3.5 \\
{} & 58478.09704 & 0.997 & 61  & -                 &  -59.1 $\pm$ 2.5 \\
\hline
TYC 2888-780-1 & 
     60215.30954 & 0.002 &  93  & -                 &   4.6 $\pm$ 11.1 \\
{} & 60215.33042 & 0.016 &  107 & -                 &   2.7 $\pm$ 11.2 \\
{} & 60215.34566 & 0.026 &  104 & -                 &   1.0 $\pm$ 8.6  \\
{} & 60215.36088 & 0.037 &  120 & -                 &  16.6 $\pm$ 13.0 \\
{} & 58470.07066 & 0.298 &  118 &  134.1 $\pm$ 9.3  & 285.4 $\pm$ 10.2 \\
{} & 58470.08685 & 0.309 &  121 & -125.6 $\pm$ 11.0 & 271.2 $\pm$ 8.3  \\
{} & 58470.10305 & 0.320 &  120 & -132.2 $\pm$ 9.3  & 264.6 $\pm$ 9.9  \\
{} & 58470.14533 & 0.349 &  116 & -122.0 $\pm$ 7.9  & 279.7 $\pm$ 7.0  \\
{} & 58470.16154 & 0.360 &  116 & -124.1 $\pm$ 10.0 & 238.7 $\pm$ 11.4 \\
{} & 58470.17775 & 0.372 &  122 & -155.2 $\pm$ 9.5  & 248.3 $\pm$ 7.7  \\
{} & 58470.19395 & 0.383 &  119 & -110.2 $\pm$ 11.1 & 245.9 $\pm$ 11.0 \\
\enddata
\end{deluxetable*}

\section{Light curve study} \label{sec:4}

In order to obtain the physical parameters of the two targets, we used the observational data from TESS Sector 59 to construct light curves, as this sector has high observation cadence and a large amount of data. After removing a few evident outliers, the observational data for the two targets are presented in Table \ref{table 3}. By combining the radial velocities and eclipse timings (see Section 5 for the calculation of eclipse timings), we conducted a joint light - radial velocity - timing analysis using the 2015 version W-D code \citep{1971ApJ...166..605W,1979ApJ...234.1054W, 1990ApJ...356..613W, 2014ApJ...780..151W} to model the light curves, radial velocities, and eclipse timings of the two eclipsing binaries.

\begin{deluxetable*}{ccccc}
\tablenum{3}
\tablecaption{TESS observation data of the two targets
\label{table 3}}
\tablewidth{0pt}
\tablehead{
\colhead{Name} & \colhead{BJD($+2400000$)} & \colhead{Phase} & \colhead{ $\mathrm{flux}_{\mathrm{SAP}}$ } & \colhead{$ \mathrm{m}_{\mathrm{TESS}} $}
}
\startdata
TYC 3740-2072-1 & 59910.26496 & 0.33090 &  1.00841 & 10.32931 \\
{}              & 59910.26728 & 0.33150 &  1.00638 & 10.33150 \\
{}              & 59910.26959 & 0.33210 &  1.00739 & 10.33041 \\
{}              & 59910.27190 & 0.33269 &  1.00816 & 10.32958 \\
{}              & 59910.27422 & 0.33329 &  1.00888 & 10.32880 \\
\enddata
\tablecomments{This table is available in its entirety in machine-readable form in the online journal.}
\end{deluxetable*}

Based on the shapes of the light curves, Mode 2 (detached binaries) was used for orbital analysis. The effective temperature provided by Gaia DR3 was used as the effective temperature of the primary component. According to the primary component's effective temperature, the gravity-darkening exponents were set as $g_1 = g_2 = 1.0$ \citep{1924MNRAS..84..665V, 1967ZA.....65...89L}, and the bolometric albedo coefficients were set as $A_1 = A_2 = 1.0$ \citep{1969AcA....19..245R}. Other coefficients were set following \citet{2007ApJ...661.1129V}. Since the secondary minima of both targets do not occur at phase 0.5, they are eccentric eclipsing binaries. The eccentricity $e_0$ and the periastron $\omega_0$ were calculated using the formulas provided by \citet{2015PASA...32...23K}:

\begin{equation}
    e_0 \cos \omega_0 = \frac{\pi}{2}[(\varphi_2-\varphi_1)-0.5],
\end{equation}

\begin{equation}
    e_0 \sin \omega_0 = \frac{w_2-w_1}{w_2+w_1},
\end{equation}
where $\varphi_1$ and $\varphi_2$ are the phases of primary and secondary eclipsing minima, respectively, and $w_1$ and $w_2$ are their respective widths. The periastron-synchronized factor $F$ is given by:

\begin{equation}
    F=\sqrt{\frac{1+e_0}{\left(1-e_0\right)^{3}}}.
\end{equation}

For TYC 3740-2072-1, $e_0=0.027(12)$, $\omega_0= 81.917(13)~deg$, and $F= 1.057(27)$; for TYC 2888-780-1, $e_0=0.039(8)$, $\omega_0= 38.244(492)~deg$, and $F= 1.084(21)$. During the analysis, the free parameters are the semi-major axis $a$, the systemic radial velocity $V_{\gamma}$, the orbital inclination $i$, the secondary surface temperature $T_2$, the mass ratio $q$, the luminosity of the primary component $L_1$, the dimensionless potentials of the primary and secondary components $\Omega_1, \Omega_2$, as well as measurable orbital period variation parameters ($t_0$, $P_0$, $\dot{P_0}$, $\dot{\omega_0}$).

\begin{figure*}
\includegraphics[width=\textwidth]{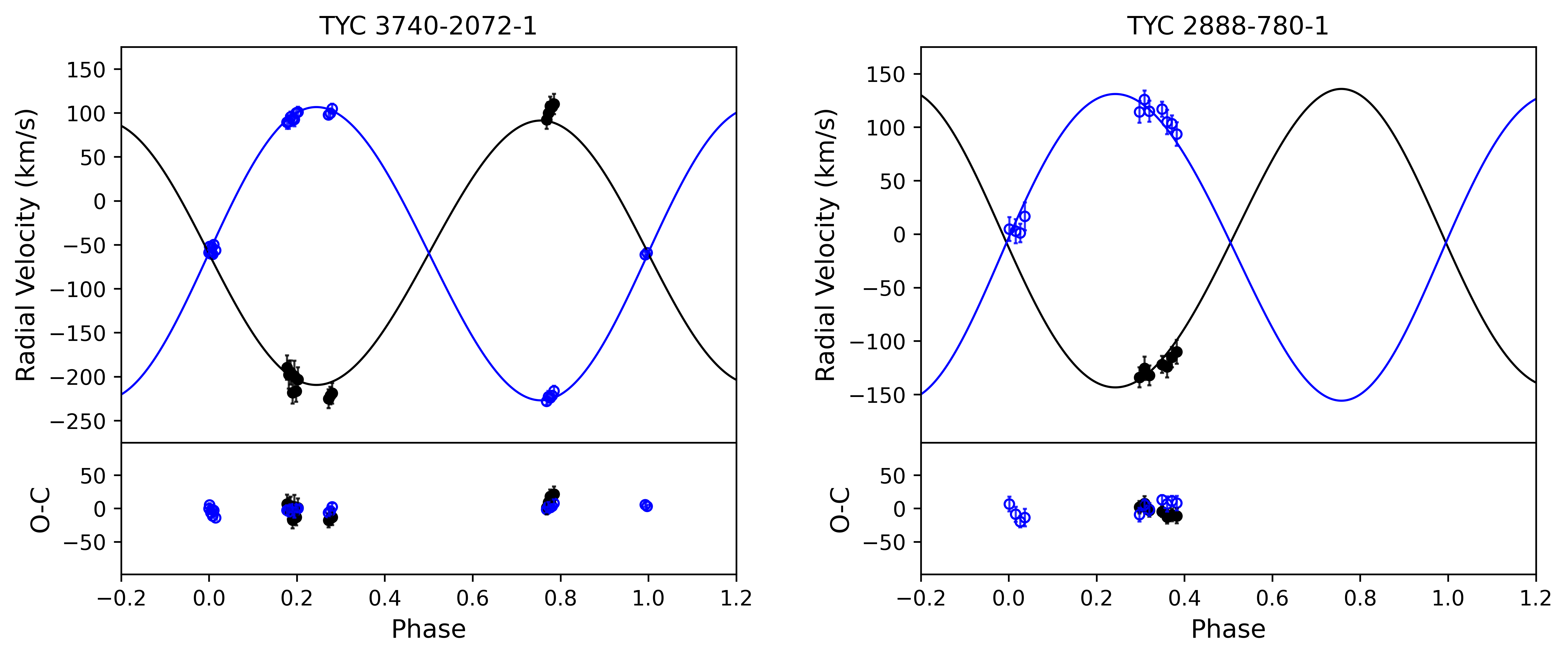}
\caption{Theoretical and observed radial velocity curves for two targets. The black line represents the theoretical curve for the primary component, while the blue line represents the theoretical curve for the secondary component. The solid circles ($\bullet$) indicate the observed results for the primary component, and the hollow circles ($\circ$) indicate the observed results for the secondary component. The lower part of the figure shows the differences between the observed values and the theoretical curves.
\label{fig:2}}
\end{figure*}

The original observational data and the fitted theoretical light curves of the two targets are shown in Figure \ref{fig:1}, while the radial velocity curves are presented in Figure \ref{fig:2}. The residuals between the observed and theoretical data are nearly flat, indicating an excellent fit of both the light and radial velocity curves. All the final determined orbital parameters and the calculated physical parameters are listed in Table \ref{table 4}.

\begin{deluxetable*}{cccccc}
\setlength{\tabcolsep}{10pt}
\tablenum{4}
\tablecaption{Physical and Orbital Period Variation Parameters of Two Systems
\label{table 4}}
\tablewidth{0pt}
\tablehead{
\colhead{} & \colhead{} & \multicolumn{2}{c}{TYC 3740-2072-1}  & \multicolumn{2}{c}{TYC 2888-780-1}                 \\
\colhead{Parameter} & \colhead{Unit} & \colhead{W-D program} & \colhead{ETA} & \colhead{W-D program} & \colhead{ETA}
}
\startdata
$T_1$           & $(K)$            & 18147                     &                        & 10902                     &\\
$T_2$           & $(K)$            & 17097      $\pm$ 15       &                        & 10341      $\pm$ 11       &\\
$V_{\gamma}$    & $(km~s^{-1})$    &  -59.58    $\pm$  1.05    &                        &    -8.05   $\pm$  2.35    &\\
$a$             & $(R_{\odot})$    &   24.472   $\pm$  0.103   &                        &     8.132  $\pm$  0.129   &\\
$i$             & $(deg)$          &   82.331   $\pm$  0.056   &                        &    88.927  $\pm$  0.114   &\\
$q (M_2/M_1)$   &                  &    0.902   $\pm$  0.003   &                        &     0.973  $\pm$  0.014   &\\
$e_0$           &                  &    0.0274  $\pm$  0.0117  &  0.0275  $\pm$ 0.0001  &     0.0397 $\pm$  0.0086  & 0.0385  $\pm$ 0.0001  \\
$t_0$           & $(BJD +2459936)$ &    
0.06034 $\pm$ 0.00014 & 0.06038 $\pm$ 0.00010 & 0.01423 $\pm$ 0.00008 & 0.01054 $\pm$ 0.00018 \\
$P_0$           & $(days)$         & 
3.86721 $\pm$ 0.00001 & 3.86722 $\pm$ 0.00002 & 1.45473 $\pm$ 0.00001 & 1.45472 $\pm$ 0.00002 \\
$\dot{P_0}$     & $(days~cycle^{-1})$ &
$(2.408 \pm 5.139) \times 10^{-9}$ &          & $(1.828 \pm 1.073) \times 10^{-9}$ &\\
$\omega_0$      & $(deg)$          &   
81.917 $\pm$ 0.013    & 82.387 $\pm$ 0.445    & 38.244 $\pm$ 0.492    & 34.527 $\pm$ 0.287    \\
$\dot{\omega_0}$ & $(deg~cycle^{-1})$ 
& 0.04115 $\pm$ 0.00009    &  0.04134 $\pm$ 0.00320 &  0.02049  $\pm$  0.00005  &  0.02019 $\pm$ 0.00029 \\
$\Omega_1$      &                  &    4.423  $\pm$  0.005  &                        &     5.906 $\pm$  0.018 &\\
$\Omega_2$      &                  &    6.462  $\pm$  0.021  &                        &     6.370 $\pm$  0.067 &\\
$M_1/M_{\odot}$ &                  &    6.914  $\pm$  0.088  &                        &     1.728 $\pm$  0.083 &\\
$M_2/M_{\odot}$ &                  &    6.233  $\pm$  0.080  &                        &     1.682 $\pm$  0.081 &\\
$R_1/R_{\odot}$ &                  &    7.113  $\pm$  0.003  &                        &     1.673 $\pm$  0.002 &\\
$R_2/R_{\odot}$ &                  &    4.105  $\pm$  0.011  &                        &     1.496 $\pm$  0.014 &\\
$L_1/L_{\odot}$ &                  & 4946.807  $\pm$  4.818  &                        &    35.651 $\pm$  0.082 &\\
$L_2/L_{\odot}$ &                  & 1298.093  $\pm$  8.193  &                        &    23.073 $\pm$  0.429 &\\
$log~g_1$       & $(cgs)$          &    3.57   $\pm$  0.01   &                        &     4.23  $\pm$  0.02  &\\
$log~g_2$       & $(cgs)$          &    4.01   $\pm$  0.01   &                        &     4.31  $\pm$  0.03  &\\
\hline
$\dot{\omega}_{rel}$ & ($deg~cycle^{-1}$)     
& 0.00123 $\pm$ 0.00001 & 0.00123 $\pm$ 0.00001 & 0.00096 $\pm$ 0.00002 & 0.00096 $\pm$ 0.00002 \\
$\dot{\omega}_{rel} / \dot{\omega}_{0}$ & $ $ 
& 0.0299       & 0.0298       & 0.0469       &0.0475        \\
$\dot{\omega}_{cl}$ & ($deg~cycle^{-1}$)      
& 0.03992 $\pm$ 0.00009 & 0.04011 $\pm$ 0.00320 & 0.01953 $\pm$ 0.00005 & 0.01923 $\pm$ 0.00029 \\
\enddata
\end{deluxetable*}

\section{Eclipse Timing Analysis} \label{sec:5}

Through eclipse timing analysis (ETA), we can investigate the evolution of binary systems (e.g., mass transfer, loss of system angular momentum) and search for potential companion objects in binary systems. We recorded the eclipsing minima from the O-C gateway \footnote{\url{http://var2.astro.cz/ocgate/}} and collected as many times of minimum light as possible to construct ETA diagrams for the two systems.

Observations of our two targets have been recorded by the Super Wide Angle Search for Planets (SuperWASP, \citealt{2010A&A...520L..10B}), All-Sky Automated Survey for SuperNovae (ASAS-SN, \citealt{2014ApJ...788...48S, 2017PASP..129j4502K, 2018MNRAS.477.3145J}), Zwicky Transient Facility survey (ZTF, \citealt{2019PASP..131a8002B, 2019PASP..131a8003M}), Kamogata/Kiso/Kyoto Wide-field Survey (KWS, \citealt{maehara_automated_2014}), and TESS. The times of minimum light calculated using the K-W method \citep{1956BAN....12..327K}, where the observational data from ASAS-SN, ZTF, and KWS are discrete, require the use of the period shift method proposed by \citet{2020AJ....159..189L} to obtain the eclipsing times. We have derived the times of minimum light for the two targets and converted them to Barycentric Julian Date (BJD) using the online tool \footnote{\url{https://astroutils.astronomy.osu.edu/time/}} provided by \citet{2010PASP..122..935E}. These values are listed in Table \ref{table 5}, where $E$ denotes the cycle number, $O-C$ represents the difference between the observed and calculated minima, and $Residuals$ indicate the deviations between the fitted curve and the O-C values.

\begin{deluxetable*}{ccccccc}
\tablenum{5}
\tablecaption{The Minimum Moment of Two Systems
\label{table 5}}
\tablewidth{0pt}
\tablehead{
\colhead{Name} & \colhead{BJD($+2400000$)} & \colhead{Error} &  \colhead{E} & \colhead{O-C} & \colhead{Residuals} & \colhead{Source}
}
\startdata
TYC 3740-2072-1 & 54382.71653 & 0.00424 &  -1436   & 0.01098 &  0.00284 & SuperWASP \\
{}              & 54388.57691 & 0.00182 &  -1434.5 & 0.07051 &  0.00018 & SuperWASP \\
{}              & 54419.51517 & 0.00143 &  -1426.5 & 0.07088 &  0.00082 & SuperWASP \\
{}              & 56917.72843 & 0.00946 &   -780.5 & 0.04988 &  0.00466 & KWS       \\
{}              & 57234.83617 & 0.00418 &   -698.5 & 0.04430 &  0.00114 & ASAS-SN   \\
{}              & 57232.86085 & 0.00262 &   -699   & 0.00259 &  0.00264 & ASAS-SN   \\
{}              & 57631.18487 & 0.00697 &   -596   & 0.00133 &  0.00032 & KWS       \\
{}              & 58816.51308 & 0.00206 &   -289.5 & 0.02181 & -0.00110 & TESS      \\
{}              & 58818.42509 & 0.00199 &   -288   & 0.00002 & -0.00004 & TESS      \\
{}              & 58820.38003 & 0.00184 &   -288.5 & 0.02152 & -0.00130 & TESS      \\
\enddata
\tablecomments{This table is available in its entirety in machine-readable form in the online journal. A portion is shown here for guidance regarding its form and content.}
\end{deluxetable*}

\begin{figure*}
\includegraphics[width=\textwidth]{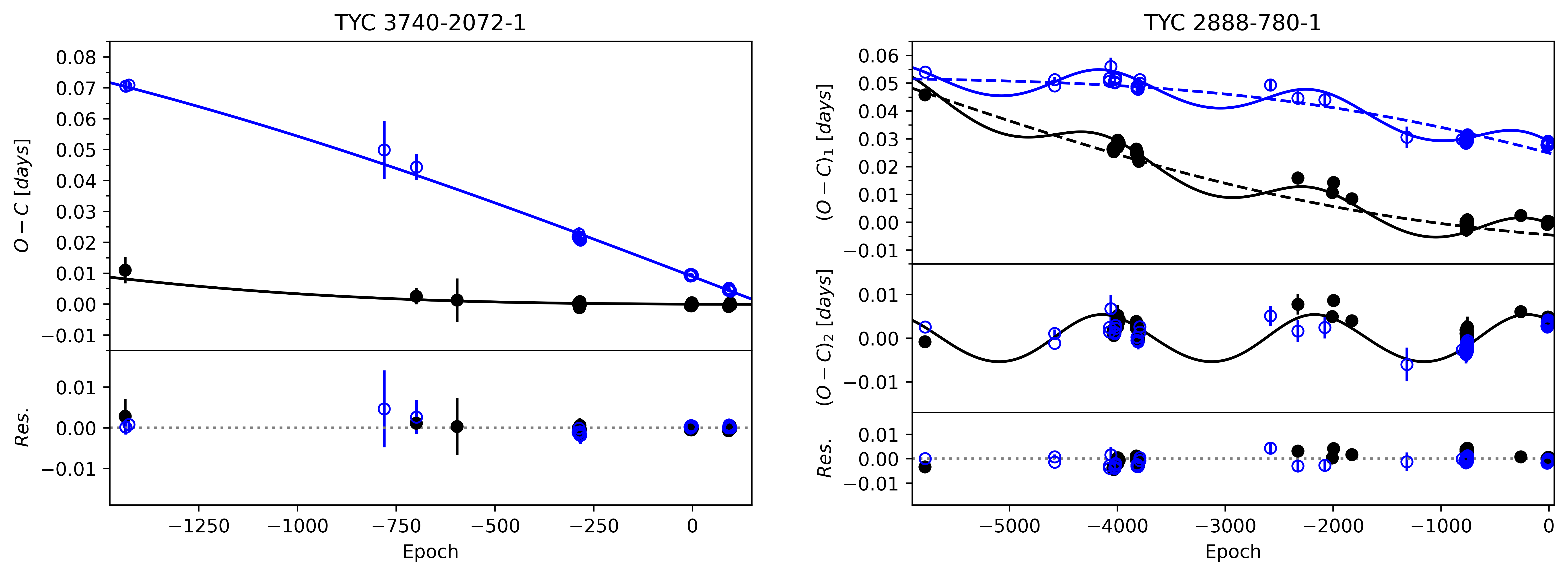}
\caption{The ETA fitting curves for two targets. Solid dots ($\bullet$) represent primary minima, while hollow dots ($\circ$) represent secondary minima. The black line represents the theoretical curve of the primary minimum, while the blue line represents the theoretical curve of the secondary minimum. The middle part of the figure on the right shows the periodic components of the ETA curve after removing the AM component.
\label{fig:3}}
\end{figure*}

We first use the linear ephemeris formula $ t = t_0 + P_0 \times E $ to calculate the ETA values for the two targets, as listed in Table \ref{table 5}. Where $t$ represents the observed time of the minimum, $t_0$ denotes the initial epoch, $P_0$ is the orbital period, and $E$ is the cycle number, as shown in Figure \ref{fig:3}. Both binaries exhibit apsidal motion (AM) \citep{1939MNRAS..99..451S}, and TYC 2888-780-1 exhibits both the AM effect and periodic variations. We use the following equations given by \citet{1995Ap&SS.226...99G} to calculate the AM parameters for both targets.

\begin{equation}
\begin{aligned}
t = & ~ t_0 + P_s \times E + \frac{P}{2\pi} \Bigg\{(2j-3) \Bigg[A_1 \cos \omega - \frac{1}{4} A_3 \cos 3\omega + \frac{1}{16}A_5 \cos 5\omega \Bigg] \\
& + \frac{1}{2}\Bigg[A_2 \sin 2\omega - \frac{1}{4} A_4 \sin 4\omega + \frac{1}{16} A_6 \sin 6\omega \Bigg] \Bigg\},
\end{aligned}
\end{equation}
where $j=1$ represents the primary minimum, $j=2$ represents the secondary minimum, and $A_i (i=1,\dots,6)$ is the function of the orbital inclination $i$ and eccentricity $e$. $P_s$ and $P$ denote the sidereal and anomalistic periods, respectively, both measured in days, and $\omega$ is the argument of periastron, which satisfies the following relations:

\begin{equation}
\omega = \omega_0 + \dot{\omega}E,
\end{equation}

\begin{equation}
P_s = P\left(1-\frac{\dot{\omega}}{2\pi}\right),
\end{equation}
where $\omega_0$ is the argument of periastron at $t_0$, $\dot{\omega}$ is the AM rate per cycle.

In the ETA curve of TYC 2888-780-1, in addition to the AM effect, periodic variations are also observed, suggesting the presence of other physical mechanisms such as magnetic activity cycles or the light-time travel effect (LTTE). Equations by \citet{1952ApJ...116..211I} can calculate the influence of periodic variations on the timing of minima:

\begin{equation}
\tau = \frac{a_{12} \sin i_3}{c} \left[\frac{1 - {e_3}^2}{1 + e_3 \cos \nu_3} \sin(\omega_3 + \nu_3) + e_3 \sin \omega_3\right],
\end{equation}

\begin{equation}
A=\frac{a_{12}\sin i_3\sqrt{1-e_3^2\cos^2\omega_3}}{c},
\end{equation}
where $a_{12}$ represent semi-major axis in astronomical units ($AU$), $e_3$, $i_3$, $\omega_3$ denote the eccentricity, inclination, and argument of periastron of the assumed third body, respectively. $c$ denotes the speed of light. $\nu_3$ is the true anomaly of the position of the eclipsing pair's mass centre on the orbit, determined by the epoch $t_3$ of periastron passage, which denotes the moment when the center of mass of the eclipsing binary reaches periastron, and the period $P_3$. $A$ represents the semi-amplitude of periodic variations, expressed in days.

We use the OCFit code provided by \citet{2019OEJV..197...71G, 2023OEJV..241....1G} to estimate the ten unknown orbital period parameters ($t_0$, $P_s$ or $P$, $e_0$, $\omega_0$, $\dot{\omega}$, $a_{12} \sin i_3$, $e_3$, $\omega_3$, $t_3$ and $P_3$). The orbital period variation parameters of the two systems are presented in Table \ref{table 4}, while the third-body parameters of TYC 2888-780-1 are listed in Table \ref{table 6}. The parameters derived using the W-D program and ETA show good agreement, demonstrating a high degree of consistency between the two methods in parameter estimation. Figure \ref{fig:3} presents the ETA diagrams for both targets.

\begin{deluxetable*}{ccc}
\tablenum{6}
\tablecaption{Third-Body Parameters of TYC 2888-780-1
\label{table 6}}
\tablehead{
\colhead{Parameter} & \colhead{Unit} & \colhead{TYC 2888-780-1}
}
\startdata
$t_3$                & ($BJD$)   & 2436469.13415 $\pm$ 665.89761 \\
$P_3$                & ($years$) & 7.848         $\pm$ 0.190     \\
$e_3$                &           & 0.0904        $\pm$ 0.0592    \\
$\omega_3$           & ($deg$)   & 56.387        $\pm$ 26.383    \\
$a_{12} \sin i_3$    & ($AU$)    & 0.935         $\pm$ 0.081     \\
$A$                  & ($days$)  & 0.00540       $\pm$ 0.00045   \\
\enddata
\end{deluxetable*}

\section{Discussion and conclusion} \label{sec:6}

\subsection{Orbital Period Analysis} \label{sec:6.1}

The observed rate of AM comprises two significant contributing components. One contribution arises from the Newtonian (classical) term, which is due to the non-spherical shape of both components, while the other is from the relativistic term, induced by general relativistic effects. The relativistic contribution can be calculated using the following equation given by \citet{1985ApJ...297..405G}:

\begin{equation}
\dot{\omega}_{rel} = 5.447 \times 10^{-4}\frac{1}{1-e^{2}} \left(\frac{M_{1} + M_{2}}{P}\right)^{2/3},
\end{equation}
where $P$ is the orbital period in days, $e$ is the orbital eccentricity, and $M_i$ are the masses of the stellar components, expressed in solar masses ($M_{\odot}$). The classical contribution to the apsidal motion rate, $\dot{\omega}_{cl}$, can be expressed as $\dot{\omega}_{cl} = \dot{\omega}_{0} - \dot{\omega}_{rel}$. Table \ref{table 4} provides the contributing components of the AM rates for the two targets.

Periodic variations can be explained as being caused by the magnetic activity of one or both components \citep{1992ApJ...385..621A} or by the LTTE effect induced by a third body \citep{2012AJ....144..136Y, 2021ApJ...922..122L}. Since the magnetic activity theory is suitable for explaining the orbital period variations of binaries containing late-type stars but is not applicable to early-type binaries \citep{2007MNRAS.380.1599Q}, we consider that its periodic variation is mainly due to LTTE caused by the presence of a third body.

If TYC 2888-780-1 is a triple system, the mass function $f(M_3)$ of the third body can be determined from the periodic variation parameters:

\begin{equation}
f(M_3)=\frac{M_3^3\sin^3i_3}{(M_1+M_2+M_3)^2}=\frac{a_{12}^3\sin^3i_3}{P_3^2},
\end{equation}
where $M_i$ is the masses of the components in $M_{\odot}$, $a_{12}$ is semi-major axis in $AU$, and $P_3$ is the orbital period in years.

Resulting in a mass function of $f(M_3) = 0.013 (4) M_\odot$. When the orbital inclination of the third body is $i=90^\circ$, the lower limit of its mass can be obtained as $M_{3,\min}=0.598(60)M_\odot$. Using the relation $M_{12} a_{12} \sin i_3 = M_3 a_3 \sin i_3$, the semi-major axis of the third body around the system's center of mass is determined to be $a_3=5.325(731)~AU$.

\subsection{Evolutionary State} \label{sec:6.2}

\begin{figure*}
\includegraphics[width=\textwidth]{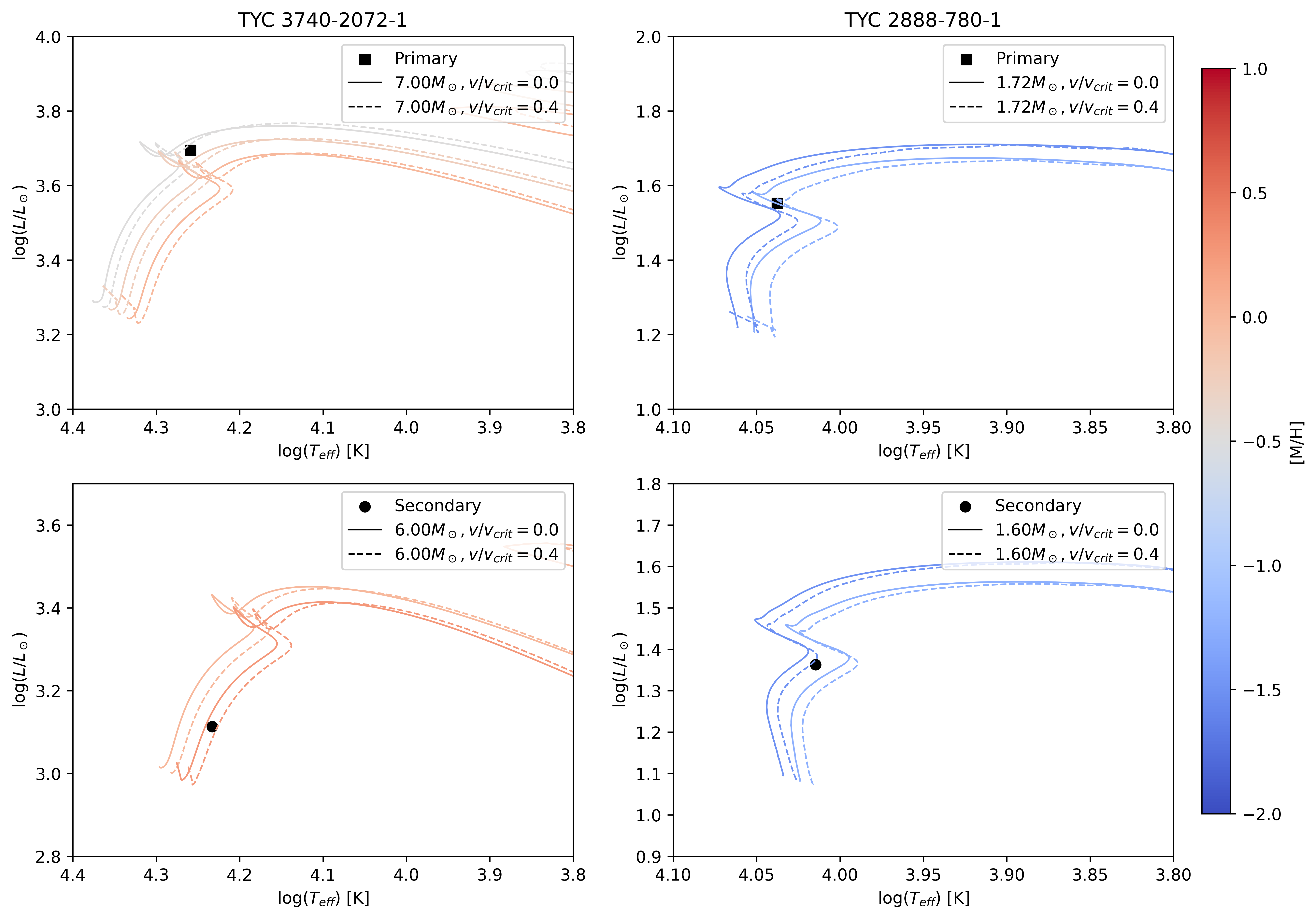}
\caption{The temperature-luminosity diagram. The solid lines represent models with a rotation rate of zero, while the dashed lines correspond to the maximum rotation rate of $v_{\rm ZAMS}/v_{\rm crit} = \omega_{\rm ZAMS}/\omega_{\rm crit} = 0.4$, where $v_{\rm crit}$ and $\omega_{\rm crit}$ are the critical surface velocity and angular velocity, respectively. The theoretical curves are provided by \citet{2016ApJS..222....8D, 2016ApJ...823..102C}.
\label{fig:4}}
\end{figure*}

Using the physical parameters obtained, we can determine the positions of the binary stars on the Hertzsprung-Russell diagram (log $T_{\mathrm{eff}}$ - log $L$), as shown in Figure \ref{fig:4}. The curves in the figure represent the theoretical evolutionary tracks calculated using the Modules for Experiments in Stellar Astrophysics (MESA) \citep{2011ApJS..192....3P, 2013ApJS..208....4P, 2015ApJS..220...15P}, corresponding to the masses of the binary components under different metallicities.

From Figure \ref{fig:4}, it can be observed that both components of TYC 2888-780-1 are in the main-sequence stage, indicating that the binary system is undergoing stable evolution. For TYC 3740-2072-1, the secondary component is in the main sequence, while the primary component has evolved beyond the main sequence and is in the subgiant phase. This is consistent with the derived surface gravity parameter of the primary component, $log~g=3.57(1)$. As the primary component continues to expand its radius, it will first fill its Roche lobe and evolve into a semi-detached binary system \citep{1955AnAp...18..379K}. Subsequently, the primary component will transfer mass to the secondary component through the Lagrangian point ($L_1$). Given the similar masses of the two components in TYC 3740-2072-1, the system may evolve into an Algol-type binary with stable mass transfer \citep{2017A&A...599A..84M}. Eventually, after losing a significant amount of mass, the primary component will evolve into a white dwarf, while the secondary component will gain mass and expand. When the secondary component fills its Roche lobe, it will transfer mass to the white dwarf. If the white dwarf's mass exceeds the Chandrasekhar limit, a thermonuclear explosion will occur, leading to a Type Ia supernova \citep{1960ApJ...132..565H, 2012NewAR..56..122W}.

In summary, through photometric and spectroscopic analyses, we obtained the physical parameters of two early type eclipsing binary systems and analyzed their orbital period variations and evolutionary states. We investigated the contributions of the AM rates for both systems and discussed the periodic variation in the orbital period of TYC 2888-780-1. We suggest that the periodic variation is caused by the LTTE effect induced by a third body, which has a minimum mass of $0.595(60)M_\odot$. Further photometric and spectroscopic observations will be required in the future to confirm the nature of this periodic variation. Additionally, we found that the primary component of TYC 3740-2072-1 has entered the subgiant phase and will evolve into a semi-detached binary system, potentially becoming a progenitor of a Type Ia supernova.

\section{acknowledgments}

The authors thank the anonymous referee for useful recommendations and suggestions that have improved the quality of the article. This work is supported by National Natural Science Foundation of China (NSFC) (No. 12273018), and the Joint Research Fund in Astronomy (No. U1931103) under cooperative agreement between the NSFC and the Chinese Academy of Sciences (CAS), and by the Qilu Young Researcher Project of Shandong University, and by the Young Data Scientist Project of the National Astronomical Data Center, and by the Cultivation Project for LAMOST Scientific Payoff and Research Achievement of CAMS-CAS. The calculations in this work were carried out at Supercomputing Center of Shandong University, Weihai.

The spectra observations were provided by the Large Sky Area Multi-Object Fiber Spectroscopic Telescope (LAMOST), which is a National Major Scientific Project built by the Chinese Academy of Sciences. Funding for the project has been provided by the National Development and Reform Commission. LAMOST is operated and managed by the National Astronomical Observatories, Chinese Academy of Sciences.

This work has made use of data from the European Space Agency (ESA) mission Gaia, processed by the Gaia Data Processing and Analysis Consortium (DPAC). Funding for the DPAC has been provided by national institutions, in particular the institutions participating in the Gaia Multilateral Agreement.

Some of the data presented in this paper were obtained from the Mikulski Archive for Space Telescopes (MAST) at the Space Telescope Science Institute. The specific observations analyzed can be accessed via \dataset[https://doi.org/10.17909/jd36-wg15]{https://doi.org/10.17909/jd36-wg15}. STScI is operated by the Association of Universities for Research in Astronomy, Inc., under NASA contract NAS5-26555. Support to MAST for these data is provided by the NASA Office of Space Science via grant NAG5-7584 and by other grants and contracts.

We acknowledge the use of data from the TESS (Transiting Exoplanet Survey Satellite), SuperWASP (Super Wide Angle Search for Planets), ASAS-SN (All-Sky Automated Survey for Supernovae), ZTF (Zwicky Transient Facility), and KWS (Kamogata/Kiso/Kyoto Wide-field Survey). These resources and efforts have been invaluable to this study and we sincerely thank the teams and organizations for their publicly available data and continuous efforts.

\bibliography{sample631}{}
\bibliographystyle{aasjournal}

\end{document}